\documentclass{article}

\usepackage{arxiv}

\usepackage[utf8]{inputenc} 
\usepackage[T1]{fontenc}    
\usepackage{hyperref}       
\usepackage{url}            
\usepackage{booktabs}       
\usepackage{nicefrac}       
\usepackage{graphicx}
\usepackage{mwe}
\graphicspath{{pictures/}}
\usepackage{amssymb}
\usepackage{amsmath}
\usepackage{color}

\title{Ultra-narrowband selective tunable filters for visible and infrared wavelength ranges}

\author{A.~D.~Utyushev\\
Siberian Federal University, Krasnoyarsk, 660041, Russia
\\Siberian State University of Science and Technology, 660014, Krasnoyarsk, Russia
\And
I.~L.~Isaev\\
Institute of Computational Modeling SB RAS, Krasnoyarsk 660036, Russia
\And
V.~S.~Gerasimov\\
Institute of Computational Modeling SB RAS, Krasnoyarsk 660036, Russia\\
Siberian Federal University, Krasnoyarsk, 660041, Russia\\
Federal Siberian Research Clinical Center under FMBA of Russia, Krasnoyarsk, 660037, Russia
\And
A.~E.~Ershov\\
Institute of Computational Modeling SB RAS, Krasnoyarsk 660036, Russia\\
Siberian Federal University, Krasnoyarsk, 660041, Russia\\
Federal Siberian Research Clinical Center under FMBA of Russia, Krasnoyarsk, 660037, Russia
\And
V.~I.~Zakomirnyi\\
Siberian Federal University, Krasnoyarsk, 660041, Russia
\\Federal Siberian Research Clinical Center under FMBA of Russia, Krasnoyarsk, 660037, Russia
\\Department of Theoretical Chemistry and Biology,\\ School of Engineering Sciences in Chemistry, Biotechnology and Health,\\Royal Institute of Technology, Stockholm, SE-10691, Sweden
\And
I.~L.~Rasskazov\\
The Institute of Optics, University of Rochester, Rochester, NY 14627, USA
\And
S.~P.~Polyutov\\
L.V. Kirensky Institute of Physics, Federal Research Center KSC SB RAS, 660036, Krasnoyarsk, Russia
\\Siberian Federal University, Krasnoyarsk, 660041, Russia
\\Federal Siberian Research Clinical Center under FMBA of Russia, Krasnoyarsk, 660037, Russia
\And
H.~\r Agren\\
Federal Siberian Research Clinical Center under FMBA of Russia, Krasnoyarsk, 660037, Russia
\\Department of Theoretical Chemistry and Biology,\\ School of Engineering Sciences in Chemistry, Biotechnology and Health,\\Royal Institute of Technology, Stockholm, SE-10691, Sweden
\And
S.~V.~Karpov\\
L.V. Kirensky Institute of Physics, Federal Research Center KSC SB RAS, 660036, Krasnoyarsk, Russia
\\Siberian Federal University, Krasnoyarsk, 660041, Russia
\\Siberian State University of Science and Technology, 660014, Krasnoyarsk, Russia
\\Federal Siberian Research Clinical Center under FMBA of Russia, Krasnoyarsk, 660037, Russia
}
\begin{document}
\maketitle
\begin{abstract}

The interaction of non-monochromatic radiation with two types of arrays comprising both plasmonic  and dielectric nanoparticles has been studied in detail. We have shown that dielectric nanoparticle arrays provide a complete selective reflection of an incident plane wave within a narrow spectral line of collective 
lattice resonance with a Q-factor of $10^3$ or larger, whereas plasmonic refractory TiN and chemically stable Au nanoparticle arrays demonstrated high-Q resonances with moderate reflectivity. The spectral position of these resonance lines is determined by the lattice period, as well as the size, shape and material composition  of the particles. Moreover, the arrays, with fixed dimensional parameters make it possible to fine-tune the position of a selected resonant spectral line by tilting the array relative to the direction of the incident radiation. These effects provide possibilities for engineering of novel  selective tunable optical high-Q filters in a wide range of wavelengths: from visible to middle IR.  Several highly refractive dielectric nanoparticle materials with low absorption are proposed for  various spectral ranges, such as  LiNbO$_3$, TiO$_2$, GaAs, Si, and Ge. 
\end{abstract}

\section{Introduction}
Design and fabrication of new compact optical elements with high-Q response in the visible, near infrared (IR), and middle IR wavelength ranges is a problem in applied optics with high priority. In this regard, much attention has been focused on devices in the form of periodic one-dimensional (1D) or two-dimensional (2D) arrays composed of plasmonic or all-dielectric nanoparticles (NPs). New ideas underlying such devices have arisen from the effect first predicted in theoretical studies of regular plasmonic structures by Schatz and Markel~\cite{Zou2004a,Zou2004,Markel2005}. According to their predictions, periodic arrays of NPs are capable to support high-Q collective lattice resonances (CLRs) in extinction spectra. CLRs occur due to the interference of fields from individual particles and the Wood-Rayleigh anomaly~\cite{Wood1902,Rayleigh1907}. In the  general case, the Wood-Rayleigh anomaly or lattice resonance arises in periodic arrays (in particular, in diffraction gratings), in which the phase of the external field of a plane wave in the vicinity of an individual array element coincides with the phase of the field produced by  neighboring elements. If this condition is satisfied within the entire array at the given wavelength, a resonance extinction occurs at this wavelength. Thus, the resonance extinction is produced by the hybrid coupling of localized low quality factor (Q-factor) resonances of NPs and their non-localized interactions covering the entire array. The position and  Q-factor of these resonances depend on the geometry of the array lattice, the material composition and the shape of the  NPs~\cite{Kravets2018}. Under particular conditions, the  Q-factor of such resonances can exceed $10^2$--$10^3$ times the Q-factor of a single NP. 

CLRs in periodic arrays of plasmonic~\cite{Auguie2008,Chu2008,Kravets2008,Ross2016,Khlopin2017,Zakomirnyi17APL,Kravets2018} and all-dielectric~\cite{Jahani2016,Liu2014c,Liu2016a,Liu2016b,Liu2018e,Wang2019a} NPs have been extensively discussed during the  recent decade owing to the  great number of  potential applications in color printing~\cite{Park2013a,Proust2016a,Dong2017b,Sun2017a}, biosensing~\cite{Enoch2004,Adato2009,Kravets2010,Bontempi2017}, lasing~\cite{Zhou2013a,Ha2018}, fluorescence enhancement~\cite{Vecchi2009} and other applications~\cite{Rajeeva2017,Wang2018g,Kravets2018}. Highly-efficient reflective filters in the form of 1D~\cite{Kodali2010b,Liu2011,Liu2014,Mazulquim2014} or 2D~\cite{Peters2010,Shen2018,Upham2018,Wang2018c,Ng2019} gratings which make it possible to extract  required spectral lines from a non-monochromatic flux represent a  most important emerging  application of CLRs.

The goal of our paper is to verify the relevance of  the concept  of CLRs for solving applied problems and the advantages they can offer. Within the frame of this goal we propose the design of a device that makes it possible to select radiation in the
reflection mode from the spectral continuum within a tunable ultra-narrow spectral line  and to control its position with a high Q-factor. To achieve this goal, the following problems are solved: obtaining data on the optimal structure of the device~--- particle size and shape, lattice period and a particle material.

\section{Methods}\label{sec:methods}
\begin{figure}
\centering
\includegraphics[width=80mm]{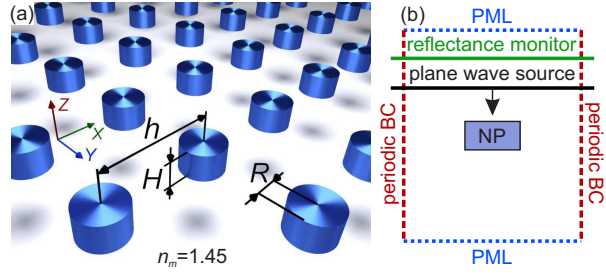}
\caption{(a)~Sketch of the ND array under consideration;  (b)~ scheme of FDTD simulation.}
\label{fig:scheme}
\end{figure}

We consider 2D arrays of NDs with height $H$ and radius $R$ arranged in a regular square lattice with period $h$, as shown in Fig.~\ref{fig:scheme}(a). The arrays are embedded in a homogeneous environment with refractive index $n_m=1.45$, which corresponds to quartz in the spectral range under study. Such structures can be fabricated using lithography technique on a quartz substrate and subsequent sputtering a layer of quartz on  top of the array. A homogeneous environment is an important factor in the model, because the  Q-factor of CLR drops dramatically in the case of the half-space geometry, where the substrate and the superstrate have different refractive indices~\cite{Auguie2010}.  The reflection spectra of such structures are calculated with commercial Finite-Difference Time-Domain (FDTD) method software~\cite{lumerical}. FDTD is a widely used computational method of electrodynamics, which in general shows excellent agreement with experimental results for CLRs~\cite{Chu2008,Thackray2015,Khlopin2017,Ng2019}. The optical response of the infinite array is simulated by considering the single particle unit cell with periodic boundary conditions (BC) applied at the lateral boundaries of the simulation box and perfectly matched layers (PML) used on the remaining top and bottom sides, as shown in Fig.~\ref{fig:scheme}(b). Arrays are illuminated from the top by  plane waves with normal incidence along the $Z$ axis and polarization along the $Y$ axis. The reflection has been calculated at the top of the simulation box using a  discrete Fourier transform monitor which is placed above the plane-wave source. The angular dependencies were obtained using the broadband fixed angle source technique~\cite{Liang2014}. An adaptive mesh has been used to accurately reproduce the nanodisk shape. Finally, extensive convergence tests for each set of parameters have been performed to avoid undesired reflections on the PMLs.

\begin{figure}
\centering
\includegraphics{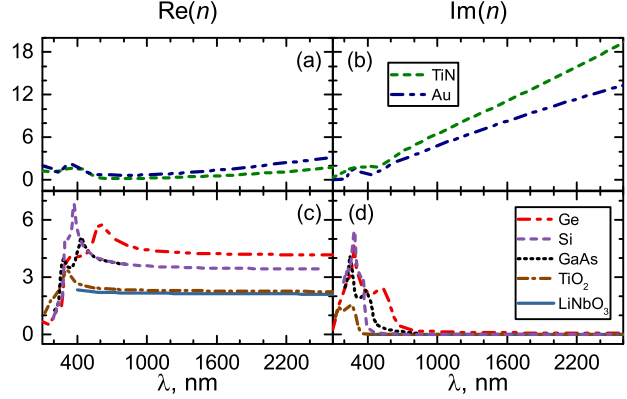}
\caption{Real and imaginary parts of complex refractive index $n$ for (a) and (b) plasmonic  TiN~\cite{Reddy2017a} and Au~\cite{Johnson1972}; (c) and (d) all-dielectric LiNbO$_3$~\cite{Zelmon1997} (here we consider only ordinary refractive index), TiO$_2$~\cite{DeVore1951}, GaAs~\cite{Aspnes1986}, Si~\cite{Edwards1997} and Ge~\cite{Potter1997}.}
\label{fig:n_k}
\end{figure}

We study arrays of both plasmonic (Au and TiN) and all-dielectric (LiNbO$_3$, Si, Ge, TiO$_2$, GaAs) nanoparticles. Figure~\ref{fig:n_k} shows tabulated experimental data for the real and imaginary parts of the  complex refractive index $n$ which have been used for each material.

\section{Results}
The lattice period $h$ varies in accordance with Rayleigh anomalies ($\pm 1$,0) and (0,$\pm 1$), the positions of which for the case of normal incidence and homogeneous environment with refractive index $n_m$ are defined as

\begin{equation}
\label{eq:lambda_xy}
\lambda_{p,q} ={h n_m}/{ \sqrt{p^2 + q^2} } \ ,
\end{equation}
where $p$ and $q$ are integers corresponding to the order of the anomaly. \eqref{eq:lambda_xy} describes the condition of constructive interference for particles within the $XOY$ plane~\cite{Bonod2016}. Note that $\lambda$ here is the vacuum wavelength. CLRs are observed in arrays of both types: plasmonic and dielectric ones. 

Before discussing CLR in NP arrays we note that the shape of the NPs is an important parameter that affects the Q-factor of the CLRs. In the calculations we examined two types of the NP shapes in the form of nanodisks and nano-parallelepipeds. Both  shapes can simply be  experimentally fabricated. 
As we found that NDs demonstrate slightly higher value of Q-factors compared to nano-parallelepipeds, the further studies were conducted with NDs only. It was found that when the disk height was equal to its radius the  Q-factor was maximum.

\subsection{Reflection spectra of plasmonic nanoparticle arrays} \label{sec:plasmonic}

Plasmonic nanoparticle arrays were the first type of structures used for observation of collective lattice resonance~\cite{Zou2004,Zou2004a,Markel2005}. Gold is a widely used material in these arrays for which  surface lattice resonances are observed in the red range of the visible spectrum~\cite{Auguie2008,Chu2008,Kravets2008}. 

\begin{figure}[htbp]
\centering
\includegraphics{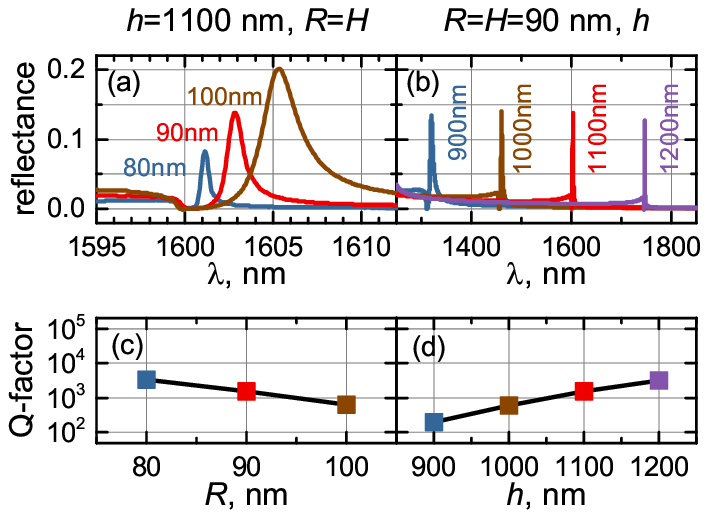}
\caption{Reflection spectra for TiN ND arrays with: (a) fixed $h=1100$~nm, and for different $R=H$ as shown inlegend, (b) fixed $R=H=90$~nm and for different $h$ as shown in legend; (c) and (d) corresponding quality factors of CLRs.} 
\label{fig:TiN}
\end{figure}

The use of TiN ND arrays (Fig~\ref{fig:TiN}) provides
 moderate reflectivity with a high Q-factor of CLRs in the telecommunication spectral range. The  optimal TiN ND radius is 90~nm. A larger particle size results in a decrease of the reflectivity and the Q-factor. This suppression of surface plasmon resonances  under extreme conditions \cite{Gerasimov16OE,Ershov2017,Gerasimov17OME} (heating of particles by pulsed laser radiation) results in  a reduction of the Q-factor and CLR amplitude. 
 
In particular, for the CLR line $\lambda=1100$~nm  in Fig~\ref{fig:TiN} the  Q-factor equals  $1.5\cdot 10^3$ at $T=23^\circ$C, $Q=1.1\cdot 10^3$ at $T=400^\circ$C, and $Q=0.7\cdot10^3$ at $T=900^\circ$C. Thus, the  high radiation resistance of TiN can be an additional advantage when using
 arrays exhibiting CLRs at high temperatures~\cite{Zakomirnyi17APL}. The use of TiN as a plasmonic material with high radiation resistance provides an extreme stability at high temperatures compared to conventional plasmonic materials (Au and Ag).

\begin{figure}[htbp]
\centering
\includegraphics{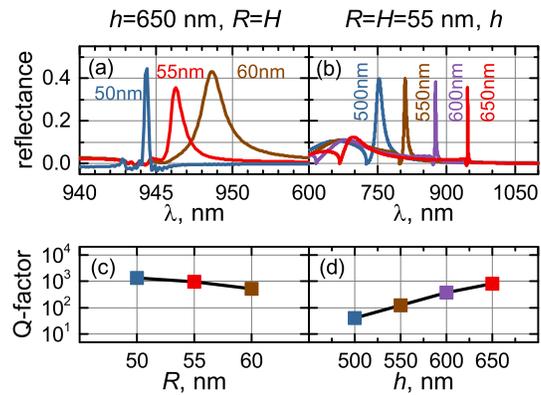}
\caption{Reflection spectra for Au ND arrays with: (a) fixed $h=650$~nm, and for different $R=H$ as shown in legend, (b) fixed $R=H=55$~nm and for different $h$ as shown in legend; (c) and (d) corresponding quality factors of CLRs.} 
\label{fig:Au}
\end{figure}

Au ND arrays (Fig~\ref{fig:Au}) demonstrate CLRs in the long-wavelength part of visible and near IR ranges~--- away from the  telecommunication range. Reflectivity and Q-factor of the Au ND array are somewhat higher compared to TiN.

\subsection{Dielectric nanoparticle arrays} \label{sec:dielectric}

\subsubsection{Reflection spectra}
In the case of dielectric structures the main requirements imposed on the type of employed NP materials are combination of high real part of the  refractive index ${\rm Re}(n)$ and a low imaginary part ${\rm Im}(n)$ for different spectral ranges which ensures low absorption inside the particles. 
 Figs~\ref{fig:n_k}(c),(d) show, in particular, the refractive indices of the following  suitable lossless materials satisfying these requirements: LiNbO$_3$, TiO$_2$, Si, GaAs, Ge. In this paper we have selected materials for each spectral range to obtain high-Q CLR in each given case.

\begin{figure}[t!]
\centering
\includegraphics{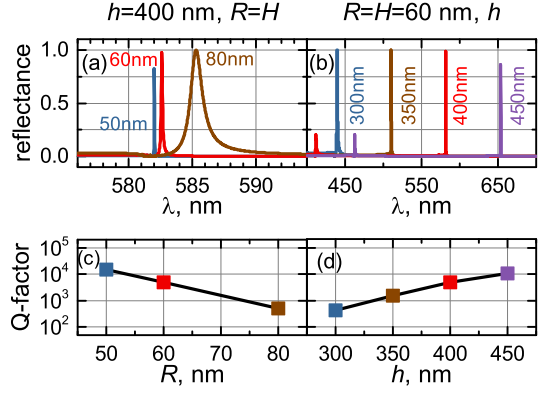}
\caption{Reflection spectra for  LiNbO$_3$ ND arrays with: (a) fixed $h=400$~nm, and for different $R=H$ as shown in legend, (b) fixed $R=H=60$~nm and for different $h$ as shown in legend; (c) and (d) corresponding quality factors of CLRs.} 
\label{fig:LiNbO3}
\end{figure}

\begin{figure}[t!]
\centering
\includegraphics{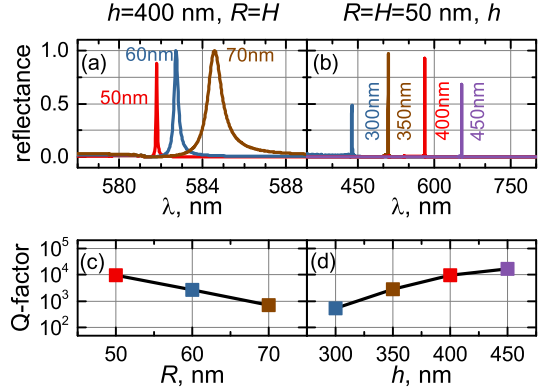}
\caption{Reflection spectra for  TiO$_2$ ND arrays with: (a) fixed $h=400$~nm, and for different $R=H$ as shown in legend, (b) fixed $R=H=50$~nm and for different $h$ as shown in legend; (c) and (d) corresponding quality factors of CLRs.}
\label{fig:TiO2}
\end{figure}

Two materials were chosen for the visible range: LiNbO$_3$ and TiO$_2$. Figures~\ref{fig:LiNbO3}(a),(b) and~\ref{fig:TiO2}(a),(b) with reflection spectra of LiNbO$_3$ and TiO$_2$ ND arrays show that the larger the particle size, the lower the Q-factor, but at the same time the higher the reflection coefficient. So the optimal combination of these factor gives a particle radius of 60~nm with a Q-factor equal to $10^3$. Figures~\ref{fig:LiNbO3}(c),(d) show that employing this material in ND arrays provides ultra-narrowband resonances in the entire visible range with a Q-factor over $10^3$ and high reflection. The utilization of TiO$_2$ in ND arrays also provides high reflectivity and Q-factor, however, with smaller size of the particle~--- 50~nm compared to LiNbO$_3$ that results in narrowing the spectral range with high reflectivity (Fig.~\ref{fig:TiO2}).  The next step is to vary the  lattice period with the given optimal radius. Calculations show that the  Q-factor of the CLR increases with wavelength. Additional opportunities to increase the CLR Q-factor are opened by changing the size and shape of the particles, as well as the optical properties of their material.

\begin{figure}[t!]
\centering
\includegraphics{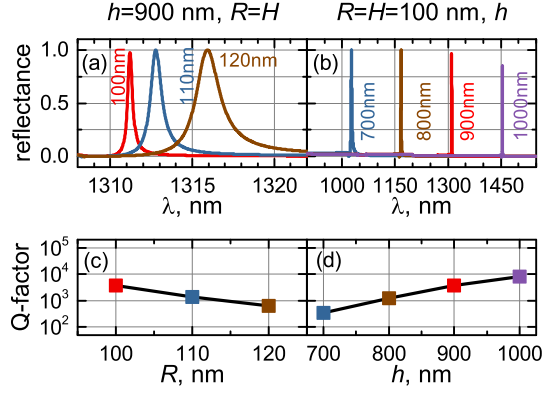}
\caption{Reflection spectra for  Si ND arrays with: (a) fixed $h=900$~nm, and for different $R=H$ as shown in legend, (b) fixed $R=H=100$~nm and for different $h$ as shown in legend; (c) and (d) corresponding quality factors of CLRs.} 
\label{fig:Si}
\end{figure}

\begin{figure}[t!]
\centering
\includegraphics{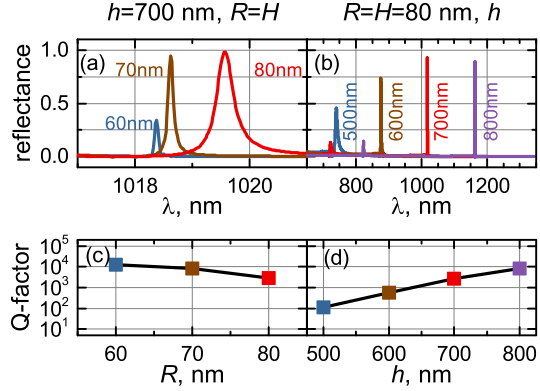}
\caption{Reflection spectra for  GaAs ND arrays with: (a) fixed $h=700$~nm, and for different $R=H$ as shown in legend, (b) fixed $R=H=80$~nm and for different $h$ as shown in legend; (c) and (d) corresponding quality factors of CLRs.}  
\label{fig:GaAs}
\end{figure}

The best materials for the near-IR range are Si and GaAs. These materials in the IR range demonstrate the properties of a dielectric (Fig.~\ref{fig:n_k}) with a near-zero imaginary part of the refractive index and its high real part, which makes it possible to excite Mie resonances in such particles~\cite{Bohren1998} with radius below 100~nm~--- much smaller than the wavelength. Fig.~\ref{fig:Si} shows the reflection spectra of Si ND arrays with ND radii $R=100$, 110, and 120~nm and radius/height ration $R/H=1$. Besides that,  Figure~\ref{fig:Si} shows that CLRs with   appear reflected radiation in the entire telecommunication wavelength range by varying the array period. Reflection spectra for GaAs ND arrays have optimal characteristics in the range between visible and telecom wavelengths (Fig.~\ref{fig:GaAs}). The optimal ND radius equals 70~nm and period 600--700~nm with ultrahigh Q-factor.  

\begin{figure}
\centering
\includegraphics{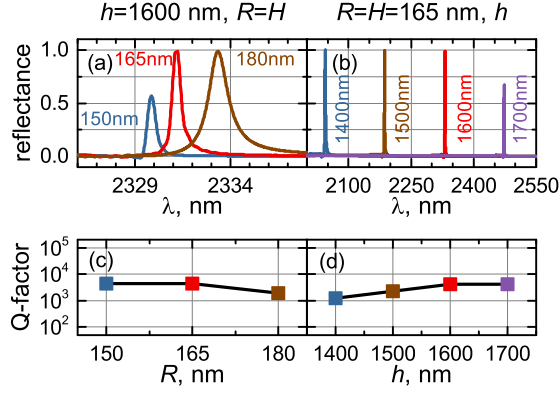}
\caption{Reflection spectra for  Ge ND arrays with: (a) fixed $h=1600$~nm, and for different $R=H$ as shown in legend, (b) fixed $R=H=165$~nm and for different $h$ as shown in legend; (c) and (d) corresponding quality factors of CLRs.} 
\label{fig:Ge}
\end{figure}

The use of the Ge  ND arrays  provides high reflectivity and Q-factor in the middle IR range with optimal ND radius 165~nm. A smaller size of the particle results in a decrease of reflectivity, larger ones are accompanied by lower Q-factors (Fig.~\ref{fig:Ge}). 

It can be seen from Figs~\ref{fig:LiNbO3}(a)--\ref{fig:Ge}(a) that, generally, the increase in the ND radius while maintaining the height $H$ is accompanied by a long-wavelength shift of the CLR and a decrease of the Q-factor. The reflection coefficient is slightly growing towards longer wavelengths (Fig.~\ref{fig:LiNbO3}--\ref{fig:Ge}) with a subsequent fall. It is found that an increase of ND height  with fixed  ratio $R/H$, results in a long-wavelength shift of the resonance and decrease of both the Q-factor and the reflection coefficient. The increase of the lattice period with preservation of the ratio $R/H$, is accompanied by a long-wavelength shift of CLR, a non-monotonic increase of the  Q-factor, and by the decrease of the reflection coefficient.

\subsubsection{Electromagnetic field configuration}
\begin{figure}
\centering
\includegraphics[width=80mm]{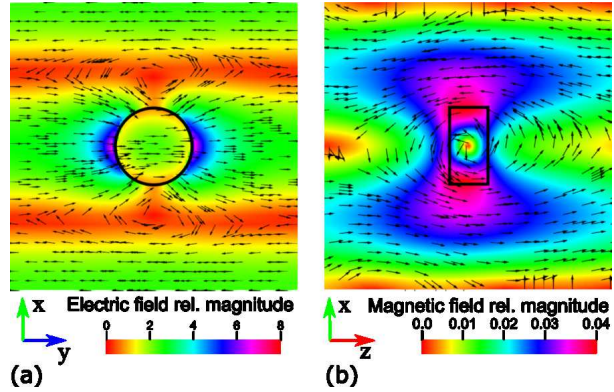}
\caption{(a) Electric and (b) magnetic field distribution in $XOY$ and $XOZ$ planes. Si ND with $R=H=120$~nm and array period $h=900$~nm at maximum reflection wavelength $\lambda=1317$~nm. Polarization direction along the $Y$ axis. Solid black line outlines the ND's boundaries.} 
\label{fig:field}
\end{figure}

The study of the radiation  reflection  from nanoparticle arrays calls for a clarification  of  what type of resonance excitation is associated with a high generated reflection.  Fig.~\ref{fig:field}(a),(b) show the configuration of electric and magnetic fields in orthogonal planes inside the unit cell  of 
Si ND with $R=H=120$~nm and array period $h=900$~nm at the maximum reflection wavelength $\lambda=1317$~nm (see the spectrum in Fig.~\ref{fig:Si}(a). Fig.~\ref{fig:field}(a) shows the spatial profile of the electric field calculated in a plane of the array, while  Fig.~\ref{fig:field}(b) shows the spatial profile of the magnetic field in the plane orthogonal to both the radiation incidence plane and the array plane. As can be seen, this configuration of the electromagnetic field coincides with the classical pattern  of an oscillating electric dipole.

\subsubsection{Dimensional invariance of collective lattice resonances}
\begin{figure}
\centering
\includegraphics{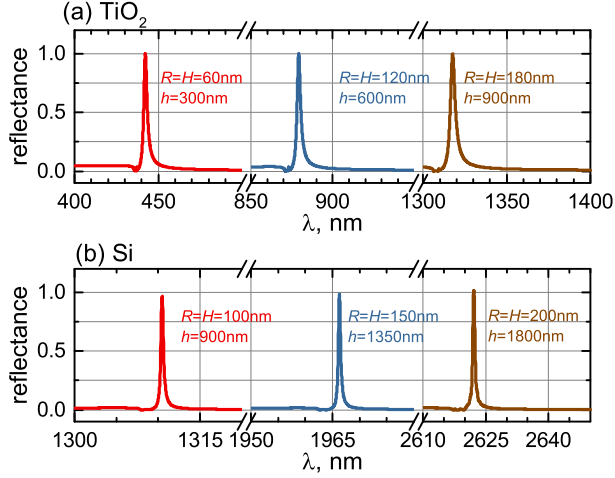}
\caption{Reflection spectra for (a) Si and (b) TiO$_2$ ND arrays with different sizes but with fixed $R/H$ and $P/R$ ratios.} 
\label{TiO2inv}
\end{figure}

Arrays of all-dielectric NPs demonstrate an important feature in conditions of utilizing  materials with zero dispersion. In particular, the  spectral position of  collective lattice resonance in TiO$_2$  NP arrays as all-dielectric systems can be predicted by multiplying all dimensional parameters of the array  (particle radius,  height and lattice period) by the same number $K$. The new resonance position will  correspond to $K\lambda_r$ (where $\lambda_r$ is the previous wavelength value (Fig.~\ref{TiO2inv}). This feature is a consequence of the scale-invariance of Maxwell's equations in the case of non-absorbing and non-dispersive materials. This is the easiest way to predict reflection at a specific wavelength. So if we determine the optimal parameters of the structure (with maximum reflectivity and the CLR Q-factor) by applying the  multiplier $K$ to all lattice parameters we can predict the  resonance position at any required  wavelength. Fig.~\ref{TiO2inv}(a),(b) show a twofold (for TiO$_2$ ND array) and one and a half (for Si ND array) increase in the parameters of the  arrays with corresponding shift of the resonance lines, something that demonstrates the scale invariance of CLRs. Scale invariance allows to design and to fabricate optical filters for operation in an arbitrary spectral range from near to far IR.  

\subsection{Fine tuning the resonance line using the angular dependence of reflection}
\begin{figure}
\centering
\includegraphics{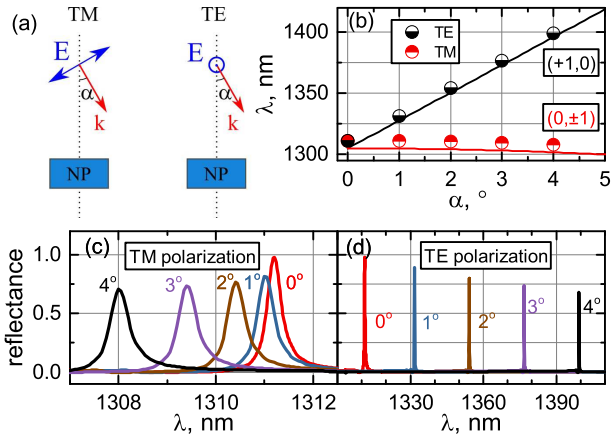}
\caption{(a) Sketch of the vector configuration for TM and TE polarization, (b) spectral positions of (+1,0) and (0,$\pm$1) Wood-Rayleigh anomalies as a functions of the incidence angle (solid lines) and corresponding positions of reflectance maxima (partially filled circles); Incidence angle dependence of the reflection spectra for Si ND array with $h=900$~nm and $R=H=100$~nm for (c) TM and (d) TE polarizations.} 
\label{fig:Si_ang}
\end{figure}

The possibility of employing NP arrays for spectral selection of radiation with required  wavelength  in the reflection mode requires the possibility to  deflect the  selected monochromatic radiation away from the incidence direction of the non-monochromatic radiation   by tilting the array at least a few degrees.  
The results from investigating this possibility are shown in Fig.~\ref{fig:Si_ang}. It is found that the slope of the array in the range of $\alpha=1^\circ-4^\circ$ is accompanied by a slight decrease in Q-factor of resonance and reflection coefficient. However, the more important effect of this tilting is the shift of the resonance line to the short-wavelength range by about 3~nm in one plane, and to the long-wavelength range by over 100~nm~--- in another plane. Obviously, the found property of arrays makes it possible
 to use them for spectral selection and to fine-tune the spectral position of the resonance line to the required wavelength.

Our calculations show that at larger angles ($\alpha>4^\circ$), the resonance shift continues to grow. However, with the increase in the angle up to $\alpha>20^\circ$ the resonance line acquires short-wavelength spectral satellites with growing amplitude, which along with a decrease of the  Q-factor impair the selectivity of filtration. To prevent such effects, one should limit the angle to $8\mbox{--}10^\circ$ or less. Similar results have been obtained for other materials.

Figs.~\ref{fig:Si_ang}(c),(d) demonstrate a decrease of the CLR Q-factor from $Q=3.6\cdot10^3$ for $\alpha=0^\circ$ to $Q=2.9\cdot10^3$ and $Q=10^3$ for $\alpha=4^\circ$ for TM and TE polarizations, respectively. Fig.~\ref{fig:Si_ang} also
demonstrates different effects of tilting the array around Y axis for different polarization of incident radiation. Rotations of the array around the $Y$ axis for TE polarization turn up to be more sensitive to  positions of the CLRs than for TM polarization. 
The explanation of this feature is given as follows. The  position  of  the Wood-Rayleigh anomaly in this case follows simple rules. The condition for constructive interference for the case of oblique incidence and square unit cell (the  wave vector is in $XOZ$ plane) reads as

\begin{equation}
k_xh = 2\pi p+kh\sin\alpha, \quad k_yh = 2\pi q \ .  
\end{equation}
Here $k_x$, $k_y$ are $x$ and $y$ components of the scattered wave vector, $p$ and $q$ are integers which represent the  the phase difference (in 2$\pi$ units) between  waves scattered by two adjacent elements of the array and incident wave  in the  $x$ and $y$ directions, $\alpha$ is the
angle between the wave vector of the incident wave and the $Z$ axis, and $k$ is the absolute value of the wave vector of the incident wave:

\begin{equation}
k = \dfrac{2\pi n_m}{\lambda} = \sqrt{k^2_x+k^2_y} \ . 
\end{equation}
For the case of $p=0$, which corresponds to the TM polarization, we have:

\begin{equation}
\lambda = \pm \frac{n_m \cos\alpha}{q} h \ .
\label{eq:anom_cos}
\end{equation}

For $q=0$, which corresponds to the TE polarization:

\begin{equation}
\lambda=\frac{n_m\left(1\pm\sin\alpha\right)}{\pm p} h \ .
\label{eq:anom_sin}
\end{equation}

Thus, the angular dependence of the spectral shift for the TM polarization obeys \eqref{eq:anom_cos}, but for TE polarization it is described by \eqref{eq:anom_sin}. These fast and slow dependencies are shown in Fig.~\ref{fig:Si_ang}(d) and do  fully correspond to the spectral features in Fig.~\ref{fig:Si_ang}(a),(b).

\section{Conclusion}
Based on the obtained results we can make the following statements. 

Periodic structures consisting of dielectric highly refractive nanoparticles with low absorption demonstrating collective lattice  resonances can be used in the reflection mode as selective  ultra-narrowband spectral filters. 
The position of spectral lines can be adjusted using the lattice period.
We have found that arrays with nanoparticles of various shapes (nanodisks and nano-parallelepipeds) demonstrate similar optical properties and can be synthesized by different available experimental techniques. 

Dielectric nanoparticle arrays are preferable structures for lossless narrowband reflection compared to plasmonic ones with  low refraction and significant absorption of highly conductive  materials in the range of the collective lattice resonances (Fig.\ref{fig:n_k}). Herewith minor array  defects, that may occur during  experimental fabrication  of the structure, do not significantly affect the Q-factor of the  CLRs~\cite{Zakomirnyi19JOSAB}.

Arrays of plasmonic nanoparticles, on the other hand, allow one to achieve a high-Q CLR response in conditions of strong NP heating (TiN), chemically aggressive media or biological environment (Au, TiN) in spite of a lower value of the reflection coefficient. The results obtained enable 
 determining the most suitable material, both taking into account its optical characteristics and operating conditions. 

Scale invariance of such arrays makes it possible to design and to fabricate filters for operation in arbitrary spectral ranges with low dispersion materials from near to far IR.

The nanoparticle arrays in the reflection mode demonstrate an optical filtering effect with  fine tuning of  the spectral position of the resonance line to the required wavelength by means of tilting the array toward the  incident radiation. 

The proposed  model provides a quantitative interpretation of the angular dependence of the characteristics of a collective lattice resonance for different geometry of the radiation incidence onto array.

\section{Funding Information}
The reported study was funded by the Russian Science Foundation (Project No.18-13-00363) (the reflection spectra of plasmonic NPs arrays); the RF Ministry of Science and Higher Education, the State contract with Siberian Federal University for scientific research in 2017--2019 (Grant No.3.8896.2017)(the reflection spectra of all-dielectric NPs arrays); A.E. thanks the grant of the President of Russian Federation (agreement 075-15-2019-676).

\bibliographystyle{unsrt}
\bibliography{Mendeley}

\end{document}